\documentclass[12pt]{article}
\usepackage{graphics} \usepackage{graphicx} \usepackage{cite}

\newcommand{\bcdot}{\mbox{\boldmath$\cdot$}}
\newcommand{\btimes}{\mbox{\boldmath$\times$}}

\begin{document}
 \title{Radiation reaction on\\ an accelerating point charge}
\author{Jerrold Franklin\footnote{Internet address: Jerry.F@TEMPLE.EDU}\\
Department of Physics\\ Temple University, Philadelphia, PA 19122}
  \date{}
   \maketitle

\begin{abstract} 
A point charge accelerating under the influence of an external force emits electromagnetic radiation that reduces the increase in its mechanical energy.  This causes a reduction in the particle's acceleration.
  We derive the decrease in acceleration due to radiation reaction  for a particle accelerating parallel to its velocity, and show that it has  a negligible effect.
   \end{abstract}

\section{Introduction}

The question of how the electromagnetic fields radiated by an accelerating charged particle produce radiation reaction that diminishes its acceleration has been of interest for a long time.
 The Abraham-Lorentz force\footnote{We are using Gaussian units with $c=1$.},
 \begin{equation}
 {\bf F_{\rm AL}}=\frac{2}{3} q^2{\bf\dot a},
 \end{equation}
 was first proposed  by Abraham\cite{a} and Lorentz\cite{l2} over 100 years ago as a retarding force. 
 A relativistic generalization of the force was formulated  by Dirac in 1938\cite{d}.
 There have a been a large number of other published and unpublished papers on 
  the Abraham-Lorentz force for many years.
  
 However, the Abraham-Lorentz force is known to
 lead to paradoxes when used in a differential equation for the motion of an accelerating charged particle.   Because of this, various modifications of the Abraham-Lorentz differential equation have been proposed. Some of these paradoxes and modifications are discussed in \cite{jdj} and \cite{dg}.
  See also\cite{w}, which includes a number of other references.

In this paper, we derive the radiation reaction on an accelerating charged particle by directly subtracting the energy radiated by the particle from its mechanical energy without introducing a radiation reaction force.  We consider the case of a point charge accelerating parallel to its velocity, which is the case generally considered in attempted derivations\footnote{There is no generally accepted derivation of the Abraham-Lorentz force.}  and applications of the Abraham-Lorentz force.

For the power radiated by an accelerating point charge, we use Larmor's formula\cite{jl}, as extended to relativity by Li\'enard\cite{li},
\begin{equation}
\frac{dW_{\rm rad}}{dt}=\frac{2}{3}q^2\gamma^6[a^2-({\bf v\btimes a})^2].
\label{prel2}
\end{equation}
For acceleration parallel to the velocity, this  reduces to
\begin{equation}
\frac{dW_{\rm rad}}{dt}=\frac{2}{3}q^2a^2\gamma^6=\frac{2}{3}q^2a'^2,
\label{prel3}
\end{equation}
where $\bf a'$ is the charged particle's acceleration in its instantaneous rest frame.

The  usual derivations for Larmor's or Li\'enard's formula
give the rate of emission of radiated energy in terms of the acceleration and velocity at a retarded time\footnote{See, for instance, Chapter 14 of\cite{jdj}
or Chapter 11 of\cite{dg}.},  
 and could not be used in a differential equation for the velocity at the present time.
In the next section we derive Li\'enard's formula for the radiated power in terms of the acceleration and velocity at the present time.
 
 \section{Electromagnetic Power Emitted\\by an Accelerating Point Charge}

 The power radiated by an accelerating point charge is given by
 the rate at which the radiated energy passes through a spherical surface of radius $R_{\rm rad}$,
 \begin{eqnarray}
P=\frac{dW_{\rm rad}}{dt}&=&\frac{1}{4\pi}\oint_S{\bf dS\bcdot[E(r},t)\btimes{\bf B(r},t)]\nonumber\\
&=&\frac{1}{4\pi}\oint {\bf{\hat r}\bcdot[E(r},t)\btimes{\bf B(r},t)]d\Omega,
   \label{Prad}
   \end{eqnarray}

    However, a major problem arises if  Eq.~(\ref{Prad})  is used to calculate the electromagnetic power. 
The fields in the integrals are to be evaluated at the present time and 
 the point $\bf r$ at which the fields are observed, but the fields for an accelerating particle are given in terms of variables given at the retarded time by the  L\'ienard-Wiechert field equations.
 
 Thus the Li\'enard formula of Eq.~(\ref{prel2}) [and Eq.~(\ref{prel3}) for acceleration parallel to the velocity] would be given in terms of the acceleration and velocity at a retarded time, $t_r$, and not at the present time, $t$.
	We show below how a radius can be chosen to give the power given by Eq.\ref{Prad}) in terms of the acceleration and velocity at the present time.

  The electric and magnetic fields appearing in Eq.~(\ref{Prad})  are given 
     by the  Li\'enard-Wiechert field equations,
\begin{eqnarray}
{\bf E(r},t)&=&\left\{\frac{q({\bf\hat r}_r-{\bf v}_r)}
{r_r^2\gamma_r^2(1-{\bf\hat r}_r\bcdot {\bf v}_r)^3}\right\}
+\left\{\frac{{\bf\hat r}_r\btimes[({\bf\hat r}_r-{\bf v}_r)\btimes{\bf a}_r]}
 {r_r(1-{\bf\hat r}_r\bcdot{\bf v}_r)^3}\right\},
\label{er1}\\ 
 {\bf B(r},t)&=&{\bf{\hat r}}_r\btimes{\bf E(r},t), \label{br1}
\end{eqnarray}
 where the variables, ${\bf r}_r,{\bf  v}_r,\gamma_r=1/\sqrt{1-v_r^2}$,
 and ${\bf a}_r$ are all  evaluated at the
retarded time, 
\begin{equation} 
t_r=t-r_r.
\label{tr}
 \end{equation}
 The radius vector, ${\bf r}_r $, is the distance from the charged particle's position at
the retarded time to the point of observation of the electromagnetic
fields at the present time.  
 
 To calculate the radiated power,
   we consider a point charge $q$ at a position ${\bf r}(t)$ with a velocity ${\bf v}(t)$ and 
   acceleration ${\bf a}(t)$.
   We make a Lorentz transformation to the rest frame of the point charge where $\bf v'=0$ and
   \begin{eqnarray}
{\bf a'_\parallel}&=&{\bf a_\parallel}\gamma^3,
\label{accl}\\
{\bf a'_\perp}&=& {\bf a_\perp}\gamma^2.
\label{acct}
\end{eqnarray}
$\bf a'_\parallel$ is the rest frame acceleration parallel to $\bf v$, and 
$\bf a'_\perp$ is the rest frame acceleration perpendicular to $\bf v$.
   
 We now evaluate the rest frame surface integral 
   \begin{eqnarray}
\frac{dW'_{\rm rad}}{dt'}&=&
\frac{1}{4\pi}\oint {\bf{\hat r}'\bcdot[E'(r'},t')\btimes{\bf B'(r'},t')]d\Omega',
   \label{Prad2}
   \end{eqnarray} 
 in the limit $R'_{\rm rad}\rightarrow 0$.
   In this limit, $t'_r=t'$ and ${\bf a'}_r={\bf a'}$
   so the electric field is given by
\begin{eqnarray}
{\bf E'(r'},t')&=&\frac{q{\bf\hat r'}}{r'^2}+\frac{q[{\bf{\hat r'}({\hat r'}\bcdot a')-a'}]}{r'}.
\label{er'}
\end{eqnarray}

Then, the surface integral in Eq.~(\ref{Prad2}) for the radiated power reduces to
\begin{eqnarray}
\frac{dW'_{\rm rad}}{dt'}
&=&\frac{q^2}{4\pi}\oint{\bf\hat r'}\bcdot[{\bf a'\btimes({\hat r'}\btimes a')}]d\Omega'\nonumber\\
&=&\frac{q^2}{4\pi}\oint[{\bf a'^2-({\hat r}\bcdot a')^2}]d\Omega'\nonumber\\
&=&\frac{2}{3}q^2a'^2.
   \label{Prad''}
   \end{eqnarray}
   The radiated power can be put back in terms of the original acceleration, using
   Eqs.~(\ref{accl}) and (\ref{acct}) to give
   \begin{eqnarray}
\frac{dW'_{\rm rad}}{dt'}
&=&\frac{2}{3}q^2(a^2_\parallel\gamma^6+a^2_\perp\gamma^4)\nonumber\\
&=&\frac{2}{3}q^2\gamma^6[a^2-({\bf v\btimes a)}^2].
   \label{Prad3}
   \end{eqnarray}
   
  The variables in Eq.~(\ref{Prad3}) are in the original moving frame, but the rate of energy emission on the left hand side of the equation is still given in terms of the rest frame variables.
However, the right-hand side of Eq.~(\ref{Prad3}) has been shown to be a Lorentz invariant\footnote{See, for instance, page 666 of \cite{jdj}.}, so Eq.~(\ref{Prad3}) can be Lorentz transformed to the moving frame, finally giving
  \begin{eqnarray}
\frac{dW_{\rm rad}}{dt}
&=&\frac{2}{3}q^2\gamma^6[a^2-({\bf v\btimes a)}^2].
   \label{Prad4}
   \end{eqnarray}
   
   This result has the same form as Li\'enard's relativistic extension of Larmor's formula, but is given here 
   with all variables at the present time, and not an arbitrary retarded time.
 
 \section{Radiation Reaction on an Accelerating Point Charge}

We want to relate $ \frac{dW_{\rm rad}}{dt}$
to its effect on the motion of an accelerating point charge.
For an accelerating particle of mass $m$, the
rate of change of its kinetic energy is given  (for $\bf a$ parallel to $\bf v$) by
\begin{eqnarray} \frac{dW_{\rm mat}}{dt}
& =&\frac{d(m\gamma-m)}{dt}=m\gamma^3{\bf (v\bcdot a)}=mva'.
  \label{mva1}
\end{eqnarray}

If an external force acts on the charged particle,  
the electromagnetic power produced by the external force
 will increase the sum of the particle's kinetic energy and the electromagnetic energy at the rates
\begin{eqnarray} 
\frac{dW_{\rm ext}}{dt}&=&\frac{dW_{\rm mat}}{dt}+\frac{dW_{\rm rad}}{dt}.
\label{www}
\end{eqnarray}
This would reduce the particle's velocity and acceleration by
\begin{eqnarray}
mva'&=&m{\bar v}{\bar a'}-\frac{2}{3}q^2a'^2,
\label{mva2}
\end{eqnarray}
where $\bar v$ and $\bar a'$ are the velocity and rest frame acceleration an uncharged partical would have.

Equation (\ref{mva2}) is a quadratic equation for $a'$, with the solution
\begin{eqnarray}
a'&=&\frac{2m{\bar v}{\bar a'}}{\left[mv+\sqrt{m^2v^2+(8/3)q^2m{\bar v}{\bar a'}}\right]},
\label{ap}\\
va'&=&\frac{2{\bar v}{\bar a'}}{\left[1+\sqrt{1+\left(\frac{q^2}{2m}\right)
\left(\frac{16{\bar v}{\bar a'}}{3v^2}\right)}\right]}.
\label{rr1}
\end{eqnarray}
Equations (\ref{mva2}), (\ref{ap}), and (\ref{rr1}) each show the decrease in the charged particle's acceleration due to the diversion of the applied energy into radiated electromagnetic energy.

We note that the increase in electromagnetic energy
 does not produce an added force on the charged particle, but is just electromagnetic energy produced in space by the action of the external force. 
  The energy going into the electromagnetic field reduces the increase of the mechanical energy of the particle.  
  
 The external force acts on two separate entities, the accelerating particle and the electromagnetic field.  The energy put into the electromagnetic field should not be considered as a separate force on the accelerating particle\footnote{A particle, even if accelerating, cannot exert force on itself.}.

To get an idea of the relative size of the radiation reaction, we consider the case of a constant external force acting on  a particle starting from rest. An uncharged particle would have a uniform acceleration ${\bar a'}$, that is constant in the particle's instantaneous rest frame.  Then we can write\footnote{This is equation (8) of \cite{rb}.}
\begin{eqnarray}{\bar a'}&=&\frac{({\bar \gamma}-1)}{x}=\frac{{\bar\gamma}^2{\bar v}^2}
{({\bar \gamma}+1)x},
\end{eqnarray}
relating $\bar a'$ to $x$, the distance traveled by the charged particle.
Then,
\begin{eqnarray}
va'&=&\frac{2{\bar v}{\bar a'}}{\left[1+\sqrt{1+\left(\frac{r_c}{x}\right)
\left[\frac{16{\bar\gamma}^2{\bar v}^3}{3v^2({\bar\gamma}+1)}\right]}\right]}.
\label{vap1}
\end{eqnarray}
We have taken the accelerating particle to be an electron, and have introduced the `classical radius' of the electron, $r_c=\frac{q^2}{2m}=2.82$ fm.

It is interesting to look at the non-relativistic and the extreme relativistic limits of   Eq.(\ref{vap1}).
The non-relativistic limit is
\begin{eqnarray}
{\bar v}<<1,\quad va'&\simeq&\frac{2{\bar v}{\bar a'}}{\left[1+\sqrt{1+\left(\frac{r_c}{x}\right)
\left[\frac{8{\bar v}^3}{3v^2}\right]}\right]},\nonumber\\
a'&\simeq&\frac{\bar a'}{\left[1+\left(\frac{2{\bar v}^3r_c}{3v^2x}\right)\right]}.
\label{nr1}
\end{eqnarray}
The relativistic limit is
\begin{eqnarray}
{\bar\gamma}>>1,\quad a'&=&\frac{\bar a'}{\left[1+\left(\frac{4{\bar \gamma} r_c}{3x}\right)\right]}.
   \label{rel}
\end{eqnarray} 

We can see from each of Eqs.~(\ref{vap1}), (\ref{nr1}), and (\ref{rel}) that radiation reaction on a charged particle, accelerating parallel to its velocity,  has a negligible effect on the  acceleration of the particle. The distance, $r_c= 2.82$ fm, is much less than any reasonable distance traveled by the particle.
  Even for a highly relativistic particle, the ratio ${\bar\gamma} r_c/x$ 
  is too small to give an observable effect\footnote{For the 50 GeV electrons at the SLAC linear collider (SLC), $\gamma r_c$ would be about 30nm.}.

\section{Conclusion}

Our conclusion is that the power radiated by an point charge, accelerating parallel to its velocity, reduces the acceleration of the particle, as shown in Eqs.~(\ref{vap1})-(\ref{rel}).  
However, the reduction of the particle's acceleration is negligible, even for highly relativistic particles.

\end{document}